\documentclass[twocolumn,showpacs,preprintnumbers,amsmath,amssymb,aps,prl]{revtex4}
\usepackage{graphicx}
\begin{document}

\title{Enhancing Mixing and Diffusion with Plastic Flow} 
\author{A. Lib{\' a}l$^{1,2}$, C. Reichhardt$^{3}$ 
and C.J. Olson Reichhardt$^{3}$ } 
\affiliation{
{$^1$}Center for Nonlinear Studies,
Los Alamos National Laboratory, Los Alamos, New Mexico 87545\\
{$^2$}Department of Physics, University of Notre Dame, Notre Dame, 
Indiana 46556\\
{$^3$}Theoretical Division,
Los Alamos National Laboratory, Los Alamos, New Mexico 87545}

\date{\today}
\begin{abstract}
We use numerical simulations to examine two-dimensional 
particle mixtures that strongly phase separate in equilibrium.
When the system is externally driven in the presence of quenched disorder,
plastic flow occurs in the form of meandering and strongly mixing channels.  
In some cases this can produce a fast and complete mixing of previously
segregated particle species, as well as an enhancement of transverse 
diffusion even in the absence of thermal fluctuations.   
We map the mixing phase diagram as a function of external 
driving and quenched disorder parameters.  
\end{abstract}
\maketitle

\vskip2pc

There have been a growing number of experiments on
collections of small particles such as colloids 
moving over periodic or complex energy landscapes generated
by various optical methods
\cite{Review,Grier,Babic,Korda,Bechinger,Spalding,Lee,Lutz}  
or structured surfaces \cite{Ling}. 
Such static and dynamical substrates can produce a variety of new 
particle segregation mechanisms \cite{Grier,Lee,Korda,Spalding} as well 
as novel types of  logic devices \cite{Babic}.  
Driven particles on 
periodic substrates can also exhibit 
enhanced diffusive properties such 
as the recently proposed giant enhancement of the diffusion
which occurs at the threshold
between pinned and sliding states
\cite{Marchesoni,Reimann,Jay,Bleil,Lacasta,Lee}.
This enhancement
has been demonstrated experimentally for colloids moving
over a periodic optical substrate \cite{Lee} 
and could be important for applications
which require mixing and dispersing of different species of particles
\cite{Lee}. 
A limiting factor for using 
diffusion enhancement 
to mix particles
is that the diffusion is enhanced 
only in the direction of the external drive. 
For instance, in a two-dimensional system with a corrugated potential that is
tilted in the 
direction of the corrugation barriers,
there is  no enhancement of the diffusion 
in the direction transverse to the corrugation barriers
at the pinned to sliding threshold. 
It would be very valuable to identify a substrate that
allows for strong enhancement of the diffusion 
in the direction transverse to the tilt of the substrate, or
one 
that would facilitate the mixing of particle
species that are intrinsically phase separated in equilibrium. 
Such a substrate could be used 
to perform fast mixing of species and 
would have applications in microfluidics, 
chemical synthesis, and creation of emulsions and dispersions.   
 
In this work we show that a phase separated 
binary assembly of interacting particles 
in the presence of a two-dimensional random substrate 
tilted by a driving field
undergoes rapid mixing and has
an enhancement of the diffusion transverse to the tilt direction.
The motion of the particles occurs via {\it plastic flow} in the
form of meandering channels which have significant excursions in the
direction perpendicular to the drive, leading to mixing of the two
particle species. 
The mixing and diffusion occur 
even in the absence of thermal fluctuations
and arise due to the complex multi-particle interactions.
We map the mixing phase diagram as a function of 
external drive and substrate properties and identify
regimes of rapid mixing. 
We find that as the difference between the two particle species increases,
the mixing becomes increasingly asymmetric 
with one species penetrating more rapidly into the other.    
Our work shows that plastic flow can be used as a mechanism 
for mixing applications, and also   
provides a new system for the study of
collective dynamical effects.

We simulate a two-dimensional system
with periodic boundary conditions in the $x$ and $y$ directions
containing two species of Yukawa particles labeled $A$ and $B$
with charges $q_A$ and $q_B$, respectively.
The particle-particle interaction potential between particles $i$ and
$j$ of charges
$q_i$ and $q_j$ at positions ${\bf r}_{i}$ and ${\bf r}_{j}$ is
$V(r_{ij}) =  E_{0}q_{i}q_{j}\exp(-\kappa r_{ij})/r_{ij}$,
where  
$E_{0} =  Z^{* 2}/4\pi\epsilon\epsilon_0$, 
$\epsilon$ is the dielectric constant, $Z^{*}$ is the unit of charge,
$\kappa$ is the screening length,  
and $r_{ij}=|{\bf r}_i-{\bf r}_j|$.
We fix $\kappa=4/a_0$ where $a_0$ is the unit of length
in the simulation.
The system size is $L=48a_0$.
The motion of particle $i$ is determined by integration of the
overdamped equation of motion
\begin{equation}
\eta \frac{d {\bf r}_{i}}{dt} = {\bf F}^{cc}_{i} + {\bf F}^{s}_{i} + 
{\bf F}_{d}
\end{equation} 
where $\eta$ is the damping term which is set equal to unity. 
Here ${\bf F}^{cc}_{i}=  -\sum^{N}_{i\neq j}{\bf \nabla} V(r_{ij})$ 
is the particle-particle interaction force,   
where $N$ is the total number of particles in the system. 
The particle density is $\rho=N/L^2$.
The substrate force 
${\bf F}^{s}_{i} = -\sum^{N_{p}}_{k=1}\nabla V_{p}(r_{ik})$
comes from
$N_p$ parabolic trapping sites placed randomly throughout the
sample.  Here
$V_{p}(r_{ik}) = -(F_{p}/2r_{p})(r_{ik} - r_{p})^2\Theta(r_{p}-r_{ik})$, 
where $F_p$ is the pinning strength, $r_p=0.2a_0$ is the pin radius,
$r_{ik}=|{\bf r}_i-{\bf r}_{k}^{(p)}|$ is the distance between 
particle $i$ and a pin at position ${\bf r}_k^{(p)}$, and $\Theta$ is
the Heaviside step function.
The pin density is $\rho_p=N_p/L^2$.
The external driving force ${\bf F}_{d}=F_d{\bf \hat{x}}$ 
is applied uniformly to all the particles.
The units of force and time are
$F_{0} = E_{0}/a_{0}$ and $\tau = \eta/E_{0}$, respectively. 
We neglect thermal fluctuations so that $T=0$. 
If the two particle species are initialized in a phase separated state,
in the absence of an external drive and disorder the particles will not
mix unless the temperature is raised above melting.

\begin{figure}
\includegraphics[width=3.5in]{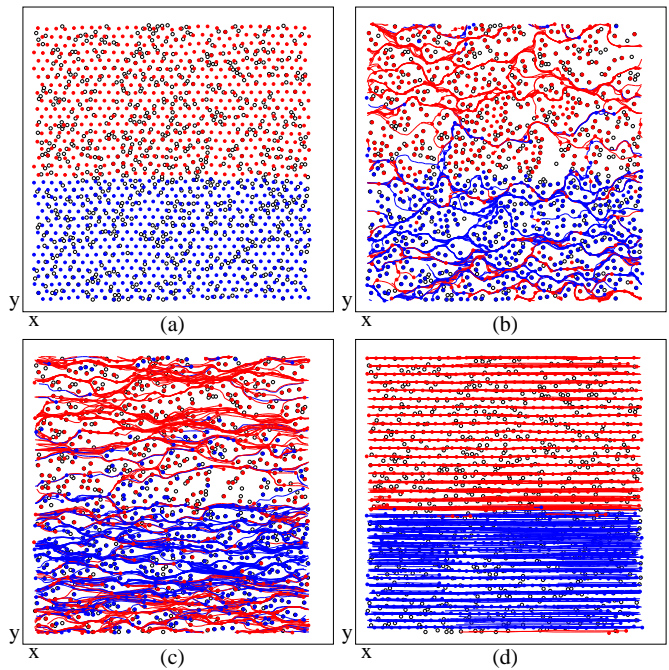}
\caption{
Red circles and red lines: particle positions and trajectories for species $A$;
blue circles and blue lines: particle positions and trajectories
for species $B$;
open black circles: pinning site locations
in a system with particle density 
$\rho=0.7$,
pin density 
$\rho_p=0.34$, and pinning force $F_p=1.0$
at different driving forces.
(a) $F_{d} = 0.0$, (b) $F_{d} = 0.1$, (c) $F_{d} = 0.4$, and (d) $F_{d} = 1.1$.
}
\end{figure}

In Fig.~1(a) we show the initial phase separated
particle configuration for 
a 50:50 mixture of the two particle species
with $q_A/q_B=3/2$ and $q_A=3$.
The particles are placed in a triangular lattice of density $\rho=0.7$ which is
immediately distorted by the pinning sites of density
$\rho_p=0.34$ and strength $F_p=1.0$.
Species $A$ occupies a larger fraction of the sample due to its larger
charge $q_A$ and correspondingly larger lattice constant
compared to species $B$.
An external driving force $F_d$ is applied in the $x$-direction 
and held at a fixed value. 

Figure~1(b) illustrates 
the particle trajectories 
at  $F_{d} = 0.1$ 
over a period of $10^5$ simulation steps. 
The trajectories form meandering riverlike structures 
with significant displacements in the direction transverse to the drive,
producing intersecting channels
that permit species $A$ to mix
with species $B$.
When the trajectories and particle positions are followed for 
a longer period of time, the 
amount of mixing in the system increases.
The riverlike channel structures 
are typical of plastic flow of particles in random 
disorder, where
a portion of the particles are temporarily trapped at pinning sites
while other particles move past, so that 
the particles do not keep their same neighbors over time.
This type of plastic flow has been observed in numerous one-component systems
including vortices in type-II superconductors 
\cite{Jensen,Dominguez,Kolton,Olson,Bassler,Higgins,Tonomura},
electron flow in metal dot arrays \cite{Middleton}, 
and general fluid flow through
random disorder \cite{Fisher,Malk}. 
These works have shown that by changing the strength and size of the
disorder, the amount of transverse wandering or 
tortuosity of the riverlike channels can be adjusted, 
and that these channels appear even for $T = 0$ 
\cite{Dominguez,Kolton,Olson,Bassler}. 
In our system we measure the diffusion in the $y$-direction, 
$d_y  = |\langle {\bf r}_i(t)\cdot{\bf \hat{y}} - {\bf r}_i(0)\cdot{\bf \hat{y}}\rangle|^2$, 
and find a long time transverse diffusive motion
with $d_y(t) \propto t^{\alpha}$ and  $\alpha = 1.0$, 
indicative of normal diffusion. 
Single component systems exhibiting plastic flow also show 
a similar transverse diffusive behavior \cite{Kolton}. 
The diffusion in our system is not induced by thermal motion but rather  
occurs due to the complex many-body particle interactions 
that give rise to the meandering riverlike channels.       
In Fig.~1(c) we plot the particle trajectories in the same system
at $F_d=0.4$.
At this drive, a larger fraction of the
particles are mobile and the riverlike channels become broader. As the
drive is further increased,
all the particles are depinned, the meandering riverlike
structures are lost, and the mixing of the particles decreases.  
Such a state is shown in Fig.~1(d) at
$F_{d} = 1.1$. 
For higher values of $F_{d}>1.1$, flow similar to that shown in 
Fig.~1(d) appears.

\begin{figure}
\includegraphics[width=3.5in]{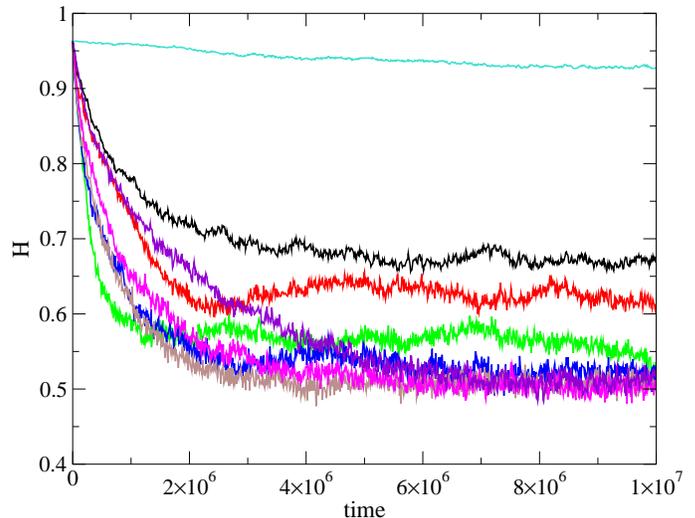}
\caption{Measure of local homogeneity $H$ vs time for the system
in Fig.~1 at $F_d=$  
0.05 (black), 0.1 (red), 0.25 (green), 0.4 (blue),
0.5 (brown), 0.6 (magenta), 0.7 (violet), and 1.1 (top curve).     
$H = 1$ for phase segregation and $H = 0.5$ for complete mixing. 
}
\end{figure}

In order to quantify the mixing, for each 
particle we identify the
closest neighboring 
particles by performing a Voronoi tesselation on
the positions of all 
particles in the system.
We then determine the probability $H$ that a 
particle is of the same
species as its neighbors.
If the system is thoroughly mixed, the local
homogeneity $H = 0.5$, while
if it is completely phase separated, $H$
is slightly less than one due to the boundary between the two 
species. 
In Fig.~2 we plot $H(t)$ for the system in Fig.~1 at
different values of $F_{d}$ 
ranging from $F_d= 0.05$
to $F_d=1.1$.
For the lower drives $F_d\le 0.1$, there are few channels and
a portion of the particles remain pinned throughout
the duration of the simulation so that 
mixing saturates near $H=0.6$ to $0.7$. 
For the intermediate drives 
$0.1 < F_d \le 0.5$ any given particle is only intermittently pinned, so
at long times all the particles 
take part in the motion and the system fully mixes, 
as indicated by the saturation of $H$ to $H = 0.5$. 
For drives $0.5 < F_{d} < 0.9$ the system can still completely mix 
but the time to reach full mixing increases with $F_d$.
At $F_{d} > 0.9$ where the particles are completely depinned, 
the mixing becomes very slow
as shown by the $H(t)$ behavior for $F_{d} = 1.1$.    
Within the strongly mixing regime, 
$H(t) \propto A\exp(-t)$ at early times before complete mixing occurs.

\begin{figure}
\includegraphics[width=3.5in]{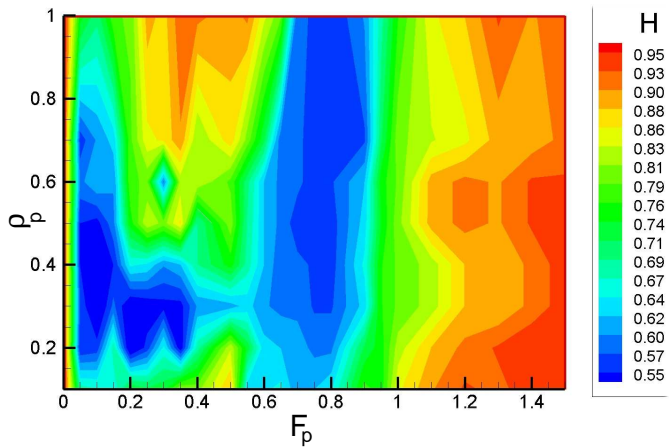}
\caption{
Mixing phase diagram of pinning density $\rho_{p}$ vs pinning strength $F_{d}$ 
in the form of a height map of the local homogeneity $H$ obtained 
from a series of simulations
with $F_{p} = 1.0$ and particle density $\rho = 0.7$. 
Strong mixing regions are blue and
weak mixing regions are red.  
}
\end{figure}

\begin{figure}
\includegraphics[width=3.3in]{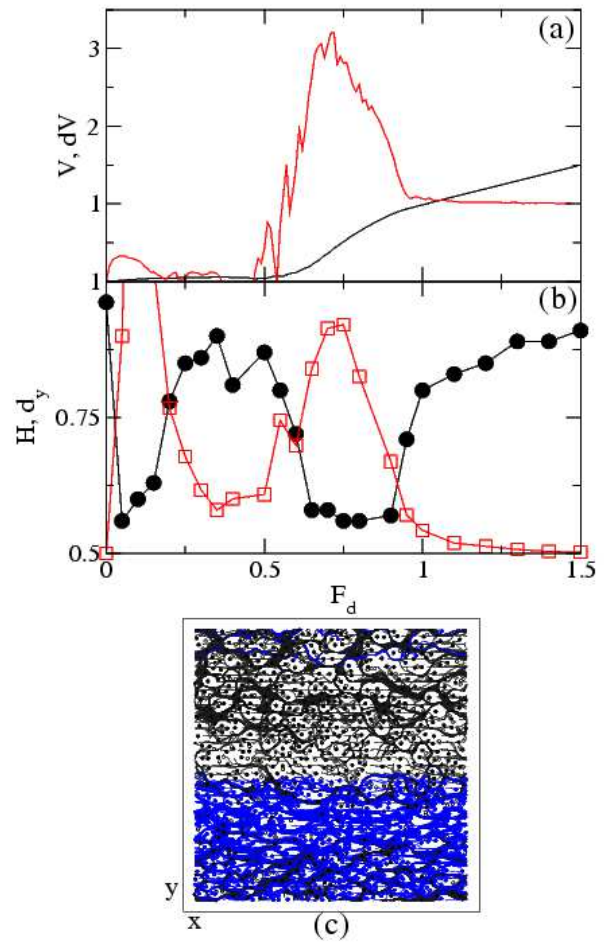}
\caption{
(a) Black line: the average particle velocity $V$ 
vs $F_{d}$ for a system with $f_{p} = 1.0$, $\rho_{p}= 0.34$, 
and $\rho = 0.7$.  Red line: the corresponding $dV/dF_d$ curve.
(b) Black circles: local homogeneity $H$; red squares: net transverse
displacement $d_y$ for the same system as in (a).
The high mixing regime ($H < 0.6$) is correlated with
enhanced transverse displacements 
and the peak in $dV/dF_d$. 
$d_y$ has been shifted down for presentation
purposes. 
(c) Particle positions (circles) and trajectories for species $A$ (black) and
species $B$ (blue) in a system with $q_A/q_B=3$
and $F_d=0.2$.
The mixing is asymmetric 
with species $A$ moving into the region occupied by species $B$ before
species $B$ moves into the area occupied by species $A$.
}
\end{figure}

In Fig.~3 we plot the mixing phase diagram of pinning density $\rho_{p}$ 
versus driving force $F_{d}$ as determined by the local homogeneity $H$
obtained from a series of simulations with $F_{p} = 1.0$ 
and $\rho = 0.7$.  The value of $H$ is measured after
$3\times 10^7$ simulation time steps. 
Blue indicates strong mixing and red indicates weak mixing.  
For $F_{d} > 1.0$ and
all values of $\rho_{p}$, all of the particles are moving in 
a fashion similar to that illustrated in Fig.~1(d). 
Since the plastic flow is lost, mixing is very inefficient
in this regime. 
For $F_d<0.6$ at high pinning densities $\rho_{p} > 0.7$, 
most of the particles are pinned, preventing a significant amount of 
mixing from occurring.
A region of strong mixing appears at  $0.6 < F_{d} < 0.9$ for all values of 
$\rho_{p}$.
Here, the particles intermittently pin and depin, producing the large amount
of plastic motion necessary to generate mixing.
There is another strong region of mixing for lower pinning densities
$0.2 < \rho_{p} < 0.4$ and low $F_{d} < 0.4$. 
In this regime there are more particles than pinning sites so that
interstitial particles, which are not trapped by pinning sites but which 
experience a caging force from neighboring pinned particles, are present.
At low drives the interstitial particles easily escape from the caging
potential and move through the system; 
however, the pinned particles remain trapped
so that the interstitial particles
form meandering paths through the pinned particles. 
This result shows that even 
a moderately small amount of disorder combined with a small drive 
can generate mixing. As the pinning
density is further decreased to
$\rho_{p} < 0.15$, the amount of mixing also decreases.   

In Fig.~4(a) we demonstrate how the mixing phases are connected to the 
transport properties of the system 
by plotting the net particle velocity 
$V=\langle N^{-1}\sum_{i=1}^N {\bf v}_i \cdot {\bf \hat{x}}\rangle$ 
and $dV/dF_d$ 
versus driving force $F_d$ for a system with $\rho_{p} = 0.34$ and
$F_{p} = 1.0$.  Here ${\bf v}_i$ is the velocity of particle $i$.
In Brownian systems, it was previously shown that
an enhanced diffusion peak is correlated with a peak in the derivative of the 
velocity force curve \cite{Marchesoni,Reimann,Jay,Bleil,Lacasta}. 
Figure 4(a) shows that there is a peak in $dV/dF_d$
spanning $ 0.5 < F_{d} < 0.9$ which also corresponds to
the region of high mixing in Fig.~3. 
There is also a smaller peak in $dV/dF_d$ 
at small drives $F_d<0.2$ produced by the
easy flow of interstitial particles. 
For $F_{d} > 1.0$, $V$ increases linearly 
with $F_d$ since the entire system is sliding freely.
In Fig.~4(b) we plot the 
local homogeneity $H$ for the same system taken from the phase diagram 
in Fig.~3.  The maximum mixing $(H<0.6)$ falls 
in the same region of $F_d$ where the peak in $dV/dF_d$ 
occurs. 
Figure 4(b) also shows that the
net traverse particle displacement $d_{y}$ 
has peaks in the strong mixing regimes.

We have also examined the effect of significantly increasing $q_{A}/q_{B}$ 
so that the system is even more strongly phase separated. 
In general, we find the same mixing features described previously; 
however, the time required for complete mixing to occur
increases with increasing $q_{A}/q_{B}$. 
The mixing also becomes {\it asymmetric}: the
more highly charged species $A$ invades the region occupied by species $B$
before the less highly charged species $B$ spreads evenly throughout the
sample.
In Fig.~4(c) we illustrate the particle trajectories 
during the first $3 \times 10^6$ simulation
time steps for a system with $q_A/q_B=3$ at $F_d=0.2$.
The mixing asymmetry can be seen from the fact that the black trails
corresponding to the motion of species $A$ overlap the blue trails representing
the motion of species $B$, but the region originally occupied by species $A$
contains no blue trails.

One issue is whether the results reported here apply more generally 
for other types of particle interactions. 
We considered only Yukawa interactions; however, the meandering channel 
structures which lead to the mixing
are a universal feature of one-component systems undergoing plastic
flow though random quenched disorder. Studies performed
on systems with long-range logarithmic interactions \cite{Kolton} 
as well as short range interactions \cite{Malk} which show this plasticity
lead us to believe that plastic flow generated by random disorder 
can produce enhanced mixing for a wide range of particle interactions. 
For our specific system of Yukawa particles, 
experiments on single component systems have
already identified a channel-like plastic flow regime \cite{Ling}. 

In summary, we have shown that 
two-dimensional plastic flow
induced by quenched disorder 
in the absence of thermal fluctuations
can lead to efficient 
mixing and enhanced diffusion in phase separating systems.
This mixing occurs due to the meandering of particles through
riverlike flow structures. 
We map the general mixing phase diagram and 
find that mixing is optimized in regimes where the particles 
depin in an intermittent fashion. For higher external drives 
the mixing is strongly reduced. These results should be general to 
a variety of 
systems where meandering flow channels appear. 

This work was carried out under the auspices of the 
NNSA of the 
U.S. DoE
at 
LANL
under Contract No.
DE-AC52-06NA25396.


\begin{thebibliography}{99}

\bibitem{Review}
D.G.~Grier, Nature (London) {\bf 424}, 810 (2003). 

\bibitem{Grier}
S.H.~Lee, K.~Ladavac, M.~Polin and D.G.~Grier, Phys.~Rev.~Lett. {\bf 94}, 110601 (2005).

\bibitem{Babic}
D.~Babic and C.~Bechinger, Phys.~Rev.~Lett.~{\bf 94}, 148303 (2005). 

\bibitem{Korda}
P.T.~Korda, G.C.~Spalding, and D.G.~Grier, Phys.~Rev.~B {\bf 66}, 
024504 (2002);
P.T.~Korda, M.B.~Taylor, and D.G.~Grier, Phys.~Rev.~Lett.
{\bf 89}, 128301 (2002). 

\bibitem{Bechinger}
M.~Brunner and C.~Bechinger, Phys.~Rev.~Lett. {\bf 88}, 248302 (2002);
K.~Mangold, P.~Leiderer, and C.~Bechinger, 
{\it ibid.} 
{\bf 90}, 158302 (2003). 

\bibitem{Spalding}
M.P.~MacDonald, G.C.~Spalding, and K.~Dholakia, 
Nature (London) {\bf 426}, 421 (2003).  

\bibitem{Lee}
S.H.~Lee and D.G.~Grier, Phys.~Rev.~Lett.~{\bf 96}, 190601 (2006).  

\bibitem{Lutz}
V.~Blickle, T.~Speck, C.~Lutz, U.~Seifert, and C.~Bechinger,
arXiv:0704.2283. 

\bibitem{Ling}
A.~Pertsinidis and X.S.~Ling, submitted. 

\bibitem{Marchesoni}
G.~Costantini and F.~Marchesoni, Europhys.~Lett.~{\bf 48}, 491 (1999).

\bibitem{Reimann}
P.~Reimann, C.~Van den Broeck, H.~Linke, P.~H\" anggi, J.M.~Rubi, and A.~Perez-Madrid,
Phys.~Rev.~E {\bf 65}, 031104 (2002). 

\bibitem{Jay}
D.~Dan and A.M.~Jayannavar, Phys.~Rev.~E {\bf 66}, 041106 (2002). 

\bibitem{Bleil}
S.~Bleil, P.~Reimann, and C.~Bechinger, Phys.~Rev.~E {\bf 75}, 031117 (2007). 

\bibitem{Lacasta}
K.~Lindenberg, A.M.~Lacasta, J.M.~Sancho, and A.H.~Romero, 
New J. Phys. {\bf 7}, 29 (2005). 

\bibitem{Jensen}
H.J.~Jensen, A.~Brass, and A.J.~Berlinsky, 
Phys.~Rev.~Lett.~{\bf 60}, 1676 (1988). 

\bibitem{Dominguez}
D.~Dom{\' \i}nguez, Phys.~Rev.~Lett~{\bf 72}, 3096 (1994). 

\bibitem{Kolton}
A.B.~Kolton, D.~Dom{\' \i}nguez, 
and N.~Gr{\o}nbech-Jensen, Phys.~Rev.~Lett.~{\bf 83}, 3061 (1999). 

\bibitem{Olson}
C.J.~Olson, C.~Reichhardt, and F.~Nori, 
Phys.~Rev.~Lett. {\bf 80}, 2197 (1998). 

\bibitem{Bassler}
K.E.~Bassler, M.~Paczuski, and G.~F.~Reiter, 
Phys.~Rev. Lett. {\bf 83}, 3956 (1999). 

\bibitem{Higgins}
S.~Bhattacharya and M.J.~Higgins, Phys.~Rev.~Lett.~{\bf 70}, 2617 (1993).

\bibitem{Tonomura}
A.~Tonomura {\it et al.}, 
Nature (London) {\bf 397}, 308 (1999). 

\bibitem{Middleton}
A.A.~Middleton and N.S.~Wingreen, Phys.~Rev.~Lett.~{\bf 71}, 3198 (1993). 

\bibitem{Fisher}
J.~Watson and D.S.~Fisher, Phys.~Rev.~B {\bf 54}, 938 (1996).

\bibitem{Malk}
M.S.~Tomassone and J.~Krim, Phys.~Rev.~E {\bf 54}, 6511 (1996); 
N.~Maleki-Jirsaraei, A. Lindner, S. Rouhani, and D. Bonn, 
J.~Phys. Cond. Mat. {\bf 17}, S1209 (2005). 

\end{thebibliography}
\end{document}